# Satellite Navigation and Control using Physics-Informed Artificial Potential Field and Sliding Mode Controller

Rakesh Kumar Sahoo[a*], Paridhi Choudhary[b] , Manoranjan Sinha[c]

rakesh.sahoo266@gmail.com[a], paridhidchoudhary@gmail.com[b], masinha@aero.iitkgp.ac.in[c]

Department of Aerospace Engineering, Indian Institute of Technology Kharagpur, West Bengal, 721302, India
*Corresponding Author

**Abstract.** Increase in the number of space exploration missions has led to the accumulation of space debris, posing risk of collision with the operational satellites. Addressing this challenge is crucial for the sustainability of space operations. To plan a safe trajectory in the presence of moving space debris, an integrated approach of artificial potential field and sliding mode controller is proposed and implemented in this paper. The relative 6-DOF kinematics and dynamics of the spacecraft is modelled in the framework of geometric mechanics with the relative configuration expressed through exponential coordinates. Various collision avoidance guidance algorithms have been proposed in the literature but the Artificial Potential Field guidance algorithm is computationally efficient and enables real-time path adjustments to avoid collision with obstacles. However, it is prone to issues such as local minima. In literature, local minima issue is typically avoided by either redefining the potential function such as adding vorticity or by employing search techniques which are computationally expensive. To address these challenges, a physics-informed APF is proposed in this paper where Hamiltonian mechanics is used instead of the traditional Newtonian mechanics-based approach. In this approach, instead of relying on attractive and repulsive forces for path planning, the Hamiltonian approach uses the potential field to define a path of minimum potential. Additionally, to track the desired trajectory planned by the guidance algorithm within a fixed-time frame, a non-singular fixed-time sliding mode controller (FTSMC) is used. The proposed fixed-time sliding surface not only ensures fixed-time convergence of system states but also guarantees the global stability of the closed-loop system without singularity. The simulation results presented support the claims made.

## 1   Introduction

With the rapid increase in space activities, the amount of space debris has grown significantly, raising the risk of collisions in orbit. Ensuring the safety and longevity of space missions now heavily depends on effective collision avoidance strategies. As a result, developing reliable and responsive guidance algorithms has become a critical need. The artificial potential field (APF) guidance algorithm is known for its computational efficiency and suitability for real-time trajectory modifications, making it effective for obstacle avoidance. However, it suffers from certain limitations—most notably the tendency to trap the satellite in local minima [1] and cause trajectory oscillations [2], particularly in regions close to obstacles or sharp potential gradients. To address these drawbacks, Several strategies have been explored in the literature. One common approach involves redesigning the potential field itself to minimize or eliminate local minima, as seen in methods like the superquadric potential field [3] and the rotational potential field [4]. Alternatively, some techniques integrate global search or optimization algorithms that help the system escape local minima, such as simulated annealing [5] or the potential gradient descent algorithm [6].

In this paper, we have proposed a novel collision avoidance algorithm physics-informed artificial potential field. The proposed Hamiltonian mechanics based artificial potential field is free from local minima. Additionally, to track the desired collision-free trajectory planned by proposed guidance algorithm within a bounded-time, a fixed-time sliding mode controller (FTSMC) is implemented in this paper. FTSMC not only ensures fixed-time convergence of the system but also guarantees global stability of the closed-loop system.

The remainder of the paper is outlined as follows: Section 2 presents the rigid body dynamic model of both target and follower satellite followed by relative translational and rotational dynamics. This is followed by proposed collision avoidance guidance algorithm presented in Section 3. Section 4 we have implemented the controller.

## 2 Rigid Body Dynamic Model of Satellites

This section presents rigid body dynamic model of target and follower satellites orbiting Earth under the influence of central gravitational field.

### 2.1 Kinematic and Dynamic Model of Target Satellite

Let the position vector of target satellite be denoted by $\tilde{b}_t \in \mathbb{R}^3$ in Inertial frame and the orientation of the target satellite be represented by rotation matrix $C_{t/i}$. Similarly, the velocity and angular velocity of target satellite is denoted by $\tilde{v}_t \in \mathbb{R}^3$ and $\tilde{\Omega}_t \in \mathbb{R}^3$, respectively. The translational as well as rotational kinematics of the target satellite can be described as

$$\dot{\tilde{p}}_t = C_{t/i}\tilde{v}_t \quad , \quad \dot{C}_{t/i} = C_{t/i}\tilde{\Omega}_t^\times \qquad (1)$$

where $\tilde{\Omega}_t^\times$ is skew-symmetric matrix form of vector $\tilde{\Omega}_t$. The translational and rotational dynamics of the target satellite can be expressed in mathematical form as

$$\begin{aligned} m_t\dot{\tilde{v}}_t &= m_t\tilde{v}_t^\times\tilde{\Omega}_t + \tilde{F}_{g/t} + \tilde{F}_{d/t} \\ J_t\dot{\tilde{\Omega}}_t &= J_t\tilde{\Omega}_t^\times\tilde{\Omega}_t + \widetilde{M}_{g/t} + \widetilde{M}_{d/t} \end{aligned} \qquad (2)$$

where $\tilde{F}_{g/t}$ and $\tilde{F}_{d/t}$ are gravity force and disturbance force, respectively; $\widetilde{M}_{g/t}$ and $\widetilde{M}_{d/t}$ are gravity gradient torque and disturbance torque, respectively acting on the target's satellite. In compact form the coupled translational and rotational dynamics of the target satellite can be written using geometric mechanics [7] as shown below.

$$\boldsymbol{P}_t\dot{\tilde{\xi}}_t = ad^*_{\tilde{\xi}_t}\boldsymbol{P}_t\tilde{\xi}_t + \tilde{\phi}_{g/t} + \tilde{\phi}_{d/t} \qquad (3)$$

where $\tilde{\phi}_{g/t}$ is the unified vector of gravitational force and gravity gradient torque vector and $\tilde{\varphi}_{d/t}$ is the unified vector of external disturbance force and torque vector acting on the target satellite; $\tilde{\xi}_t = \left[\tilde{\Omega}_t, \tilde{v}_t\right]^T$ is unified velocity vector of translational and rotational velocity; and $\boldsymbol{P}_t$ is unified matrix of inertia and mass of target satellite.

### 2.2 Kinematic and Dynamic Model of Follower Satellite

The coupled translational and rotational dynamics of the follower satellite can be written using geometric mechanics [7] as shown below.

$$\boldsymbol{P}_f\dot{\tilde{\xi}}_f = ad^*_{\tilde{\xi}_f}\boldsymbol{P}_f\tilde{\xi}_f + \tilde{\phi}_{g/f} + \tilde{\phi}_{d/f} + \tilde{\phi}_{c/f} + \tilde{\phi}_{APF} \qquad (4)$$

where $\tilde{\phi}_{g/f}$ is the unified vector of gravitational force and gravity gradient torque vector and $\tilde{\varphi}_{d/f}$ is the unified vector of external disturbance force and torque vector acting on the follower satellite; $\tilde{\xi}_f = \left[\widetilde{\Omega}_f, \tilde{v}_f\right]^T$ is the unified velocity vector of translational and rotational velocity; and $\boldsymbol{P}_f$ is unified matrix of inertia and mass of follower satellite.

## 2.3 Relative Kinematics and Dynamics of Follower w.r.t Target Satellite

The relative configuration of follower satellite w.r.t target satellite is represented by $\tilde{\rho} = \left[\tilde{\gamma}, \tilde{b}\right]^T$ where $\tilde{\gamma}$ and $\tilde{b}$ are relative orientation and position of follower satellite w.r.t the target satellite. The relative velocity of the follower satellite with respect to the target satellite in the follower's body-fixed frame can be mathematically described as

$$\tilde{\xi} = \tilde{\xi}_f - Ad_{H^{-1}}\tilde{\xi}_t \qquad (5)$$

where $\tilde{\xi} = [\widetilde{\Omega}_r, \tilde{v}_r]$ is the unified relative velocity vector of rotational and translational velocity. The relative acceleration of follower satellite with respect to the target satellite is described as

$$\dot{\tilde{\xi}} = \dot{\tilde{\xi}}_f + ad_{\tilde{\xi}}Ad_{H^{-1}}\tilde{\xi}_t - Ad_{H^{-1}}\dot{\tilde{\xi}}_t \qquad (6)$$

The relative dynamics of follower satellite with respect to the target satellite is given by

$$\boldsymbol{P}_f\dot{\tilde{\xi}} = \tilde{\phi}_{g/f} + \tilde{\phi}_{d/f} + \tilde{\phi}_{c/f} + \tilde{\phi}_{APF} + \boldsymbol{P}_f(ad^*_{\tilde{\xi}_f}\mathbb{I}_f\tilde{\xi}_f + ad_{\tilde{\xi}}Ad_{H^{-1}}\tilde{\xi}_t - Ad_{H^{-1}}\dot{\tilde{\xi}}_t) \qquad (7)$$

## 3 Collision-Free Path Planning Algorithm of Follower Satellite

This section presents a novel collision avoidance algorithm called Artificial Potential Field (APF) based on Hamiltonian mechanics for a follower satellite operating in an obstacle-prone environment, ensuring robustness by eliminating local minima issues.

### 3.1 Artificial Potential Field path planning algorithm

Artificial Potential Field is a widely used collision avoidance path planning algorithm where the target satellite is modeled as negatively charged particle generating attractive force whereas follower satellite and obstacles are modeled as positively charged particle generating repulsive force. The attractive potential function guiding the follower satellite towards the target satellite is given by

$$\widetilde{U}_a = \tfrac{1}{2}\mu_a\big(\tilde{b}\big)^2 \qquad (8)$$

where $\mu_a$ is a positive constant and $\tilde{b}$ is the relative position of the follower satellite with respect to the target satellite. Similarly, the repulsive potential function guiding the follower satellite away from obstacles is given by

$$\widetilde{U}_r = \tfrac{1}{2}\mu_r\left(\tfrac{1}{\tilde{b}_0}\right)^2 \qquad (9)$$

where $\mu_r$ is a positive constant and $\tilde{b}_0$ is the relative position of the follower satellite with respect to an obstacle. The resultant force acting on the follower satellite is the sum of the gradient of the attractive and repulsive potential function

$$\tilde{F}_{APF} = \nabla \tilde{U}_a + \nabla \tilde{U}_r = \mu_a \tilde{b} - \mu_r \left(\frac{1}{\tilde{b}_0}\right)^3 \tag{10}$$

APF is computationally efficient path planning algorithm but suffers from local minima issue where attractive and repulsive force balances out each other i.e., $\mu_a \tilde{b} = \mu_r \left(\frac{1}{\tilde{b}_0}\right)^3$. At this point net force acting on the satellite $\tilde{F}_{APF}$ becomes zero and the satellite get trapped in local minima.

### 3.2  Physics-Informed Artificial Potential Field Path Planning Algorithm

Conventional Artificial Potential Field (APF), based on Newtonian mechanics, suffers from local minima as the potential function depends only on relative position. To overcome this, we have proposed physics-informed APF based on Hamiltonian mechanics. Unlike the traditional approach, the proposed approach incorporates both relative position and relative velocity, thus eliminating local minima. The potential function dependent on relative position is referred to as static potential, whereas the velocity-dependent function is termed kinetic potential, enabling a more dynamic and robust navigation strategy.

#### 3.2.1  Attractive Artificial Potential Field

The expression of attractive static potential as a quadratic function of the relative position of the follower satellite with respect to the target satellite is given by

$$\tilde{U}_a = \frac{1}{2}\mu_a (\tilde{b})^2 \tag{11}$$

where $\mu_a$ is a positive constant and $\tilde{b}$ is the relative position of the follower satellite with respect to the target satellite. Similarly, the expression of attractive kinetic potential as a quadratic function of the relative velocity of the follower satellite with respect to the target satellite is given by

$$\tilde{K}_a = \frac{1}{2}\mu_a (\tilde{v}_r)^2 \tag{12}$$

where $\tilde{v}_r$ is the relative velocity of the follower satellite with respect to the target satellite. The attractive Hamiltonian function is defined as the sum of kinetic and static potential as shown below.

$$\begin{aligned}\tilde{H}_a &= \tilde{K}_a + \tilde{U}_a = \frac{1}{2}\mu_a (\tilde{v}_r)^2 + \frac{1}{2}\mu_a (\tilde{b})^2 \\ \tilde{H}_a &= \frac{1}{2\mu_a}(\tilde{p}_a)^2 + \frac{1}{2}\mu_a (\tilde{b})^2\end{aligned} \tag{13}$$

where $\tilde{p}_a$ is the attractive momentum factor defined as $\tilde{p}_a = \mu_a \tilde{v}_r$. The attractive force $\tilde{F}_a$ acting on the follower satellite can be computed by solving the Hamiltonian equation as shown below.

$$\dot{\tilde{b}} = \frac{\partial \tilde{H}_a}{\partial \tilde{p}_a} , \quad \dot{\tilde{p}}_a + \frac{\partial \tilde{H}_a}{\partial \tilde{b}} = \tilde{F}_a \tag{14}$$

The attractive force acting on the follower satellite based on the proposed approach is a function of relative position $\tilde{b}$ and relative acceleration $\dot{\tilde{v}}_r$ as shown below.

$$\tilde{F}_a = \mu_a \dot{\tilde{v}}_r + \mu_a \tilde{b} \tag{15}$$

#### 3.2.2  Repulsive Artificial Potential Field

The expression of repulsive static potential as a quadratic function of the relative position of the follower satellite with respect to the obstacle is given by

$$\tilde{U}_r = \frac{1}{2}\mu_r \left(\frac{1}{\tilde{b}_o}\right)^2 \tag{16}$$

Similarly, the expression of repulsive kinetic potential as a quadratic function of the relative velocity of the follower satellite with respect to the obstacle is given by

$$\widetilde{K}_r = \frac{1}{2}\mu_r \left(\frac{1}{\tilde{v}_o}\right)^2 \tag{17}$$

The repulsive Hamiltonian function is defined as the sum of kinetic and static potential as shown below.

$$\begin{aligned}\widetilde{H}_r &= \widetilde{K}_r + \widetilde{U}_r = \frac{1}{2}\mu_r \left(\frac{1}{\tilde{v}_o}\right)^2 + \frac{1}{2}\mu_r \left(\frac{1}{\tilde{b}_o}\right)^2 \\ \widetilde{H}_r &= \frac{1}{2\mu_r}(\tilde{p}_r)^2 + \frac{1}{2}\mu_r\left(\frac{1}{\tilde{b}_o}\right)^2\end{aligned} \tag{18}$$

where $\tilde{p}_r$ is the repulsive momentum factor defined as $\tilde{p}_r = \frac{\mu_r}{\tilde{v}_o}$. The repulsive force $\widetilde{F}_r$ acting on the follower satellite can be computed by solving the Hamiltonian equation as shown below.

$$\dot{\tilde{b}}_o = \frac{\partial \widetilde{H}_r}{\partial \tilde{p}_r} \quad , \quad \dot{\tilde{p}}_r + \frac{\partial \widetilde{H}_r}{\partial \tilde{b}_o} = \widetilde{F}_r \tag{19}$$

The repulsive force acting on the follower satellite based on the proposed approach is a function of relative position $\tilde{b}_o$ and relative acceleration $\ddot{\tilde{v}}_o$ as shown below.

$$\widetilde{F}_r = -\frac{\mu_r \ddot{\tilde{v}}_o}{(\tilde{v}_o)^2} - \frac{\mu_r}{(\tilde{b}_o)^3} \tag{20}$$

The resultant force acting on the follower satellite due to target and surrounding obstacles is computed as the sum of attractive and repulsive forces. This can be mathematically expressed as:

$$\widetilde{F}_{APF} = \widetilde{F}_a + \widetilde{F}_r = \mu_a \tilde{b} - \frac{\mu_r}{(\tilde{b}_o)^3} + \mu_a \ddot{\tilde{v}}_r - \frac{\mu_r \ddot{\tilde{v}}_o}{(\tilde{v}_o)^2} \tag{21}$$

It can be inferred by comparing the expression of collision-avoidance force acting on the follower satellite due to conventional APF and the proposed physics-informed APF as shown in Eq. 10 and Eq. 21, respectively that the proposed approach includes two additional terms dependent on relative velocity and relative acceleration. Specifically, while the conventional APF only accounts for position-based attractive and repulsive forces, the proposed method introduces dynamic terms that depend on the relative velocity and time derivatives of the relative velocity. As a result, even when the net force due to relative position becomes zero leading to local minima issue in conventional APF, the presence of velocity and acceleration-dependent terms ensures a non-zero net force. Thus, eliminating the local minima issue.

## 4 Design approach of Fixed-Time Sliding Mode Controller

The fixed-time sliding surface consisting of the relative configuration error $\tilde{\rho}$ and relative velocity error $\tilde{v}_r$ is defined as

$$\tilde{s} = \tilde{\xi} + \mu_1 sig^{k_1}(\tilde{\rho}) + \mu_2 sig^{k_2}(\tilde{\rho}) \tag{22}$$

where $k_1, k_2 > 0$, $0 < l_1 \leq 1, l_2 > 1$. The fixed-time reaching phase based control law to ensure that the system states to reach the sliding surface within fixed time is as shown below.

$$\dot{\tilde{s}} = -\mu_{s1} sig^{k_1}(\tilde{s}) - \mu_{s2} sig^{k_2}(\tilde{s}) \tag{23}$$

The control input for fixed-time convergence of the system is given by [7]

$$\begin{aligned}\tilde{\phi}_{c/f} = &-\tilde{\phi}_{g/f} - \tilde{\phi}_{d/f} - \boldsymbol{P}_f \left(ad^*_{\tilde{\xi}_f} \boldsymbol{P}_f \tilde{\xi}_f + ad_{\tilde{\xi}} Ad_{\boldsymbol{H}^{-1}} \tilde{\xi}_t - Ad_{\boldsymbol{H}^{-1}} \dot{\tilde{\xi}}_t \right. \\ &\left. + (\mathbb{Q}_1 + \mathbb{Q}_2) \boldsymbol{G}(\tilde{\rho}) \tilde{\xi}\right) - \mu_{s1} sig^{k_1}(\tilde{s}) - \mu_{s2} sig^{k_2}(\tilde{s})\end{aligned} \tag{24}$$

Consider a continuous differentiable Lyapunov function: $V_1 = \frac{1}{2}\tilde{s}^T \boldsymbol{P}_f \tilde{s}$. The time derivative of the Lyapunov function is expressed as $\dot{V}_1 = \tilde{s}^T \boldsymbol{P}_f \dot{\tilde{s}}$

$$\begin{aligned}\dot{V}_1 &\leq -\mu_{s1}\left(\sum_{i=1}^{6}|s_i|^2\right)^{\frac{k_1+1}{2}} - \mu_{s2} 6^{\frac{l_2-1}{2}}\left(\sum_{i=1}^{6}|s_i|^2\right)^{\frac{k_2+1}{2}} \\ \dot{V}_1 &\leq -k_{t1} V^{\frac{l_1+1}{2}} - k_{t2} V^{\frac{l_2+1}{2}}\end{aligned} \tag{25}$$

Therefore, the system states will reach the desired sliding surface $\tilde{s} = 0$ given by within fixed-time, $T_{max}$ as shown below.

$$T(x) \leq T_{max} = \frac{2}{k_{t1}(1-l_1)} + \frac{2}{k_{t2}(l_2-1)} \quad (26)$$

The closed loop feedback system is globally fixed-time stable based on the control input.

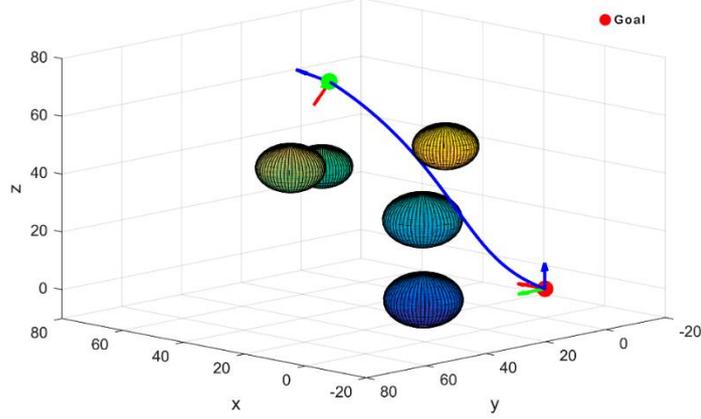

*Figure 1: Illustration of relative 3D trajectory of chaser satellite with respect to target satellite in presence of surrounding obstacles*

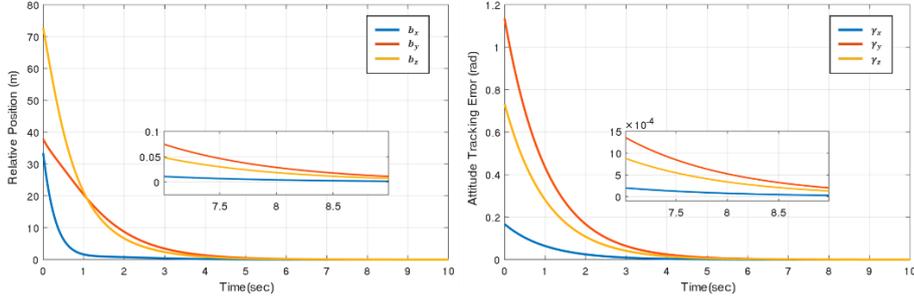

*Figure 2: Plot of relative position and orientation of follower satellite w.r.t target satellite*

## 5    Results and Discussion

The target satellite is assumed to be in Molniya orbit whose eccentricity is 0.72. The mass of both target and follower satellite is taken as 110 kg. The relative 3D trajectory of follower satellite with respect to target satellite along with coordinate frame at the beginning and end of the simulation is shown in Figure 1. The plot of relative position and orientation is shown in Figure 2 and it can be observed that the upper bound of relative position and relative orientation is $0.1 m$ and $15 \times 10^{-4} rad$ respectively. Similarly, it can be observed that the upper bound of relative translational and angular velocity is $0.04 m/s$ and $15 \times 10^{-4} rad$/s respectively.

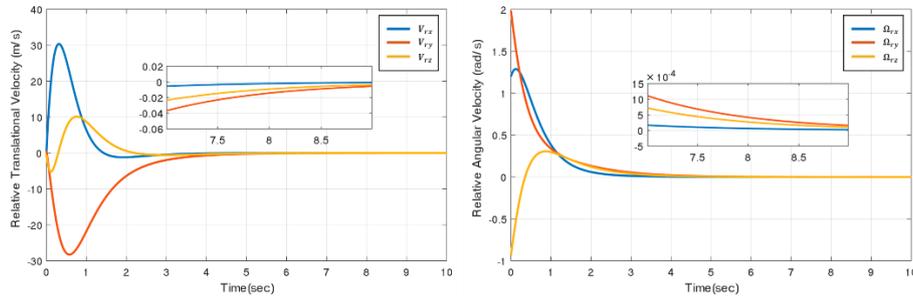

*Figure 3: Plot of relative translational and angular velocity of follower w.r.t target satellite*

## 6  Conclusion

In this paper, we have proposed a novel collision avoidance algorithm physics-informed artificial potential field which is free from local minima. Moreover, fixed-time sliding mode controller was implemented to track the desired docking configuration within fixed-time.